%% file: ngmass7-root-english.tex
\documentclass[a4paper,12pt]{article}
\usepackage{amsmath,amsfonts,amssymb}
\usepackage[T2A]{fontenc}                    
\usepackage[utf8x]{inputenc}                 
\usepackage[english]{babel}       
\usepackage{xspace}
\usepackage{url}
\usepackage{footnote}
\usepackage{nicefrac}
\usepackage{cancel}
\usepackage[uline]{hhtensor}

\usepackage{setspace}

\usepackage[draft]{prelim2e}

\usepackage{topcapt}
\usepackage[footnotesize,centerlast]{caption}
\usepackage{graphicx}
\usepackage{color}
\usepackage{wrapfig}
\graphicspath{{images/}}

\def\G     {\ensuremath{\mathbf{G}}\xspace}
\def\SIkg  {\ensuremath{\, \textit{kg}\,}}
\def\SIm   {\ensuremath{\, \textit{m}\,}}
\def\SIrho {\ensuremath{\, \textit{kg}/\textit{m}^3 }}
\def\SIG   {\ensuremath{\, \textit{m}^3/(\textit{kg}\cdot\!\textit{s}^2) }}

\begin{document}

\vspace*{-40mm}

\begin{center}
	\begin{minipage}{138mm}
		\centering
        \textbf{\Large Modeling the Evolution of a cluster of gravitating bodies
        	taking into account their absolutely inelastic collisions}
	\end{minipage}
\end{center}
\bigskip

\centerline{Kiryan~D.\,G., Kiryan~G.\,V.}

\begin{center}
\begin{minipage}{110mm}
	\emph{\small Institute of Problems of Mechanical Engineering of RAS\\
	61 Bolshoy Prospect V.O., 199178, Saint Petersburg, Russia\\
	e-mail: \textit{diki.ipme@gmail.com}}
\end{minipage}
\end{center}

\vspace*{-10mm}

\input{ngmass7-abstract-english}

\input{ngmass7-subject-english}
\input{ngmass7-problem-english}
\input{ngmass7-initial-conditions-english}
\input{ngmass7-beta-english}
\input{ngmass7-computer-simulation-english}

\input{ngmass7-conclusion-english}

\bibliographystyle{ieeetr}
\bibliography{%
    ../../../../___General-LATEX-BIB/Kiryan-bibliography-general-utf8,%
    ../../../../___General-LATEX-BIB/Kiryan-bibliography-prive-utf8}

\end{document}

%% file: ngmass7-abstract-english.tex
%

\section*{}
\label{section:ngmass7-abstract-english}

\begin{spacing}{0.80}
\begin{center}
\begin{minipage}[t]{1\textwidth}
	{\small Numerical simulation of evolution of a cluster of a finite number of gravitating bodies interacting only by their intrinsic gravity has been carried out. The goal of the study was to reveal the main characteristic phases of the spatial distribution of material bodies constituting the cluster.  In solving the problem, the possibility of interbody collisions was taken into account, the collisions being assumed to be absolutely inelastic. Forces external to the body cluster under consideration were ignored. Among all the internal force factors acting within the cluster, only the gravitational interaction was taken into account. The total mass of all the gravitating bodies of the cluster was assumed to remain constant during the entire evolution. The Cauchy problem with natural initial conditions was considered. To check the process of solution, the so-called rotation curve was used which represents the current radial distribution of orbital velocities of the cluster bodies. The numerical analysis showed time variations of the model cluster rotation curve and, particularly, the fact that the rotation curve horizontal section is only a short moment in evolution of the gravitating bodies cluster. The results obtained within the scope of classical mechanics show that it is possible to represent all the rotation curve variations for the observed galaxies without appealing to the hypothesis of non-observable gravitating “dark matter”.}
\end{minipage}
\end{center}
\textbf{Key words}: galaxy rotation curve, $n$-body problem, evolution of gravitating masses, 
								evolution number, “dark matter”.
\end{spacing}


%% file: ngmass7-subject-english.tex
%

\section{Object of the study}
\label{section:ngmass7-subject-english}

\begin{figure}[h!]
	\centering
	\includegraphics[scale=1.0]{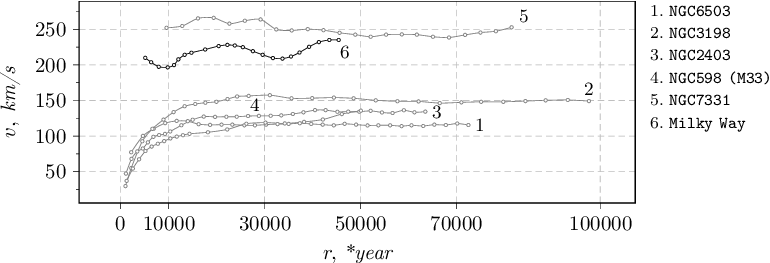}
	\caption{%
		Examples of galaxy rotation curves from~\cite{2017:IJAFR:Evolution}.
		$v$~is the radial velocity,\\ $r$~is the distance expressed in time units}
	\label{fig:rotation7-fig-Galaxies-velocity}
\end{figure}

The object of this study is the galaxy rotation curve that represents orbital velocity distribution of the stars over their distances from the current galaxy center of mass. Rotation curves of some galaxies contain plateaus, namely, sections where the constancy of star orbital velocities is observed%
\footnote{%
	Based on systematic observations of the 21th spiral galaxy 
	(i.e., measurements of Doppler shifts of the star spectral lines)~\cite{1980ApJ...238..471R}, 
	V.C.~Rubin has obtained a characteristic radial distribution of the orbital velocity containing a section looking like a plateau.%
}.
Fig.~\ref{fig:rotation7-fig-Galaxies-velocity} presents rotation curves of some of the observable galaxies having characteristic plateaus.
Just the difficulties with physical interpretation of the plateau existence in rotation curves of some galaxies gave rise to such a hypothetical invisible gravitating entity as “dark matter” whose destiny was to explain the observed constancy of radial velocities.


%% file: ngmass7-problem-english.tex
%

\section{Problem definition}
\label{section:ngmass7-problem}

Our task was to show through a classical-mechanics numerical experiment that the characteristic shape of the rotation curve of the gravitating bodies cluster may be simulated without involving additional gravitating matter in any possible form.  Evolution of clusters of gravitating bodies is controlled by a number of factors of various physical natures~\cite{1976:book:Alfven:Evolution}. This is a process of continuous interaction between all the material components of the cluster accompanied by spatial redistribution of the gravitating bodies and change in their number. Herewith, the accompanying accretion processes cause the mass concentration, i.e., matter compaction around the cluster's instantaneous center of mass.

Let us consider dynamics of a closed system consisting of $n$~bodies (homogeneous spheres) with
masses $m_i,i=1,\dots,n$ taking into consideration only absolutely inelastic collisions. We do not consider kinematic restrictions on the cluster components. All the force factors external to the system are excluded. Internal force interactions between the bodies are limited to only the gravitational interaction, non-gravitational factors are ignored. The body-to-body collisions are assumed to be absolutely inelastic. 

\begin{figure}[ht!]
	\centering
	\includegraphics[scale=1.0]{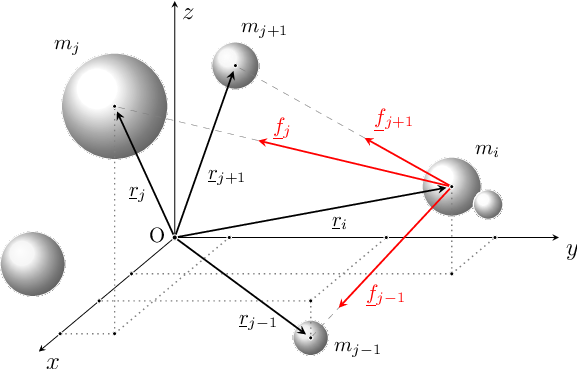}
	\caption{%
		Schematic diagram of a gravitational impact upon a cluster body with mass~$m_i$ from bodies with masses $m_{j-1}$, $m_{j}$ and $m_{j+1}$%
	}
	\label{fig:ngmass7-fig-forces}
\end{figure}

Let us consider a fixed Cartesian frame of reference $Oxyz$ (Fig.~\ref{fig:ngmass7-fig-forces}) 
where spatial location and velocity of each $i$-th body of the cluster are defined by radius-vectors~$\vec{r}_i$ and $\vec{v}_i$, respectively. Each body interacts gravitationally with all the others. Gravity field of each material body is characterized by the field intensity vector~$\vec{E}\:$:
\begin{equation}\label{eq:ngmass7:E}
	\vec{E}_{ji} =
	\G\:
	\frac{m_j}{\:|\vec{r}_j\!-\vec{r}_i|^2\:} \cdot 
	\frac{\vec{r}_j\!-\vec{r}_i}{\: |\vec{r}_j\!-\vec{r}_i| \:}\;,
\end{equation}
where $\vec{E}_{ji}$ is the gravity field intensity vector of the $j$-th body at distance $\vec{r}_{j}-\vec{r}_{i}$, and \scalebox{0.9}{$\;\G\!=\!6.673 84(80)\!\times\! 10^{-11}\SIG$} is the matching scale-dimension factor%
\footnote{%
	Notice that in analyzing the data presented by International Committee on Data for Science and Technology (CODATA) it is possible to speak about only two decimals in the SI system. This statement was justified in~\cite{2018:IJAFR:GConst}.%
}.
Due to additivity, we can sum up gravitational forces acting upon body~$m_i$ from all the other bodies of the system. A set of second-order differential equations modeling dynamics of gravitational interaction between the $n$~bodies looks as follows:
\begin{equation}\label{eq:ngmass7:system}
	m_i\:\frac{d^2\vec{r}_i}{dt^2}=
		\sum\limits^n_{ j=1,\;  j \neq i }
		\overbrace{\: \vec{E}_{ji}\;m_i\:}^{\vec{f}_j} + \;
		\cancelto{0}{\vec{f}^\ast(t,\vec{r}_i,\dot{\vec{r}}_i)\phantom{\Big.}}\;,
		\qquad i=1,\dots,n\;.
\end{equation}
Here $\vec{f}^\ast$ is the sum of all forces of other than gravitational nature acting on moving body~$m_i$ from its material environment. This term was introduced only for formal completion of the expression. In the scope of our task those forces are excluded from consideration.

Our task was not only to calculate the model cluster bodies' dynamics taking into account gravitational accretion but also to construct a rotation curve for each time point of the numerical experiment.

In the cluster under study, the accretion mechanism materializes via absolutely inelastic collisions between the cluster bodies. To our opinion, the absolutely inelastic collision is that: the new (integrated) body formed due to the contact of two bodies continues moving in the cluster gravity field with the velocity dictated by the principle of momentum conservation. For instance, in the case of collision of two bodies with masses and velocities $m_1$, $m_2$ and $\vec{v}_1$, $\vec{v}_2$, respectively, velocity of the “stuck” bodies is
\begin{equation}\label{eq:ngmass7-v}
	\vec{v}=\frac{\: m_1\:\vec{v}_1 + m_2\:\vec{v}_2 \:}{m_1+m_2}\;.
\end{equation}

Thereafter, to solve the set of second-order differential equations~\eqref{eq:ngmass7:system} it is necessary to clear up the question of natural initial conditions.


%% file: ngmass7-initial-conditions-english.tex
%

\section{Natural initial conditions}
\label{section:ngmass7-initial-conditions}

Now let us speak about initial conditions for the problem. Since we are going to simulate really observable evolution phases of the gravitating bodies forming a cluster (e.g., a galaxy), the initial conditions should comply with the real galaxy state at a chosen time point. This is just what we call natural initial conditions.

Not going into details, we can assume that, taking into account gravitational accretion, evolution of gravitating body clusters may be conditionally divided into three main stages:
\begin{enumerate}
	\item
		\underline{Initial}: spatial distribution of the gravitating matter with material inclusions
		whose density is higher due to primary accretions of various natures.
		\vspace*{-2mm}
	\item
		\underline{Borderline}: clearly pronounced ordered rotation of the cluster bodies about its instantaneous center of mass with a distinct plateau in the cluster rotation curve.
		\vspace*{-2mm}
	\item
		\underline{Final}: almost all the matter is maximally concentrated near the center of rotation, manifestations of the gravitational accretion are minimal.
\end{enumerate}

Based on the above, initial conditions for the defined task (see formula~\eqref{eq:ngmass7:system}) were assumed to be coordinates and velocities of the cluster bodies  at the moment when the cluster rotation curve has not yet a pronounced plateau, i.e., at the initial stage of the evolution.

Assume that initially all the cluster bodies move along circular orbits. However, there are questions about what is the mutual arrangement of body orbits and what values of linear velocities should be the preset. To answer those questions, it is necessary to define the natural initial conditions corresponding to the evolution stage preceding the borderline state so as to pass in the course of the numerical experiment through the borderline state to the final evolution phase.

In paper~\cite{2017:IJAFR:Evolution} it was shown that the initial distribution of orbit radii of the cluster bodies may be expressed as follows:
\begin{equation}\label{eq:ngmass7-alpha-r}
	r_i = r_{min} \left( \sqrt[3]{i} \right)^\alpha\!,\quad i=1,\ldots n\;,
\end{equation}
where $r_i$ is the circular orbit radius of the $i$-th body of the system; $r_{min}$ is the minimal radius of the circular orbit; $\alpha$ is the parameter defining the character of the initial distribution of cluster body circular orbits. 

To remain at the circular orbit $r_i$ in radius, the $i$-th body should move with an appropriate orbital velocity~$v_i$ to be calculated as
\begin{equation}\label{eq:ngmass7-vm}
	v_i=\sqrt{\G\frac{\: m_i^s \:}{r_i}}\;, \quad
	m_i^s=m^\ast + \sum\limits_{j=1}^{i-1}m_j\;, \quad i=1,\ldots n\;,
\end{equation}
where $v_i$ is the orbital velocity of the $i$-th body with mass~$m_i$;
$m^s_i$ is the sum of masses of all the bodies in the volume of the \textit{Sphere}~$r_i$ in radius; $m^\ast$ is the mass around which body~$m_1$ rotates along the circular orbit~$r_1$ in radius.

\begin{figure}[htb!]
	\centering
	\includegraphics[scale=1.1]{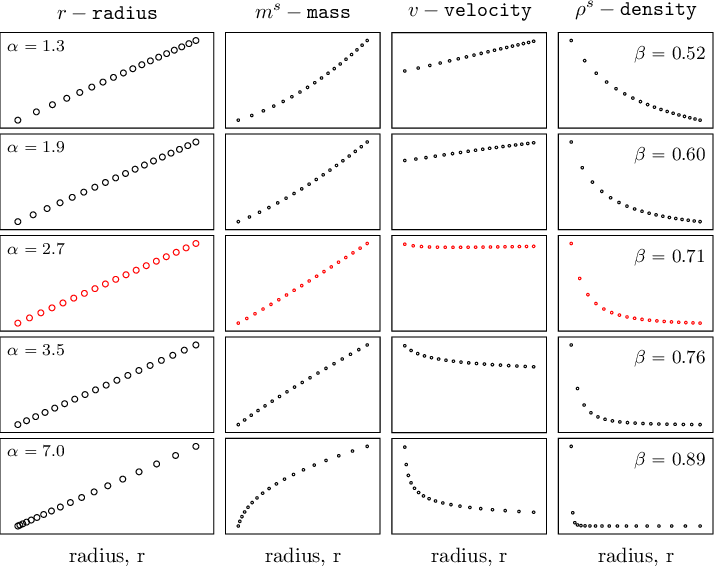}
	\caption{%
		Variants of the natural initial values of the  body orbit radii at different moments of the cluster evolution. $\beta$~is the evolution number~\cite{2017:IJAFR:Evolution}%
	}
	\label{fig:ngmass7-fig-InitialOrbites}
\end{figure}

Fig.~\ref{fig:ngmass7-fig-InitialOrbites} demonstrates the distribution of the cluster body orbits calculated via~\eqref{eq:ngmass7-alpha-r} and~\eqref{eq:ngmass7-vm} at different~$\alpha$.

Let us choose according to Fig.~\ref{fig:ngmass7-fig-InitialOrbites} such a value of $\alpha$ at which the gravitating bodies comprising the cluster are in the state preceding the borderline phase when the rotation curve contains a plateau, namely,~$\alpha\!<\!2.7\:$. Thus, we have defined natural initial conditions necessary to continue solving the task in hand~\eqref{eq:ngmass7:system}. 

Direction of the cluster bodies rotation about the cluster instantaneous center of mass may be chosen arbitrary. It does not affect the mechanism of evolution.


%% file: ngmass7-beta-english.tex
%

\section{Numerical estimation of the current\\ evolution stage}
\label{section:ngmass7-beta}

What should be the procedure for quantitative evaluation of the state of the process involving continuous variations in the spatial locations of all the gravitating bodies and in their number and masses?

Let us use the method of nested \textit{Spheres} whose principles were described in detail in paper~\cite{2017:IJAFR:Evolution}. This method enables calculation of the evolution number~$\beta$ characterizing the current radial distribution of the gravitating bodies forming the cluster. Evolution number~$\beta$ ranges from~$0$ (the initial evolution phase) to $1$ (the final evolution phase) and is invariant with respect to the cluster size and masses of bodies comprising it. To~determine the current (design or observable) body coordinates, let us construct an ordered (from minimum to maximum radius) sequence of nested~\textit{Spheres} whose radii correspond to the distances between the gravitating bodies and the cluster's instantaneous center of mass.

Evolution number~$\beta$ is a factor in power function
\begin{equation}\label{eq:ngmass7-rho-beta}
	\rho(r) = \rho_{min} \left(\frac{r}{r_{max}}\right)^{-3\beta}\;,\qquad
	\rho_{min} > 0\:,\quad
	0 \leqslant \beta \leqslant 1\:.
\end{equation}
Here
$r$ is the \textit{Sphere} radius;
$\rho$ is the \textit{Sphere} density;
$\beta$ is the dimensionless factor (evolution number);
$r_{max}$ is the orbit radius of the outermost body (star) of the considered cluster;
$\rho_{min}$ is the density of the \textit{Sphere} with radius~$r_{max}$.

To calculate evolution number~$\beta$ via formulae~\eqref{eq:ngmass7-alpha-r} and \eqref{eq:ngmass7-vm}, let us form a sequence of averaged \textit{Sphere} densities
\begin{equation}\label{eq:ngmass7-rho}
	\rho_i^s=m_i^s \! \biggm/ \! \frac{4}{3}\pi r_i^3\;,\quad i=1,\ldots,n\;,
\end{equation}
where $\rho^s_i$ is the averaged density of the $i$-th \textit{Sphere}.

Evolution number~$\beta$ is a result of approximation of the obtained density distribution~\eqref{eq:ngmass7-rho} by the power function~\eqref{eq:ngmass7-rho-beta}. Approximation by other analytical functions gives essentially worse results. The approximation quality was estimated via correlation factor~$\mathcal{R}$.


%% file: ngmass7-computer-simulation-english.tex
%

\section{Numerical simulation}
\label{section:ngmass7-computer-simulation}

As a computational model of the gravitating bodies cluster, $n$~homogeneous spheres of one and the same mass and density were taken.
\begin{equation}\label{eq:ngmass7-m-rho}
	m_i=100\SIkg\;,\quad
	\rho_i=2500\SIrho\;,\quad i=1,\ldots,n=100\;.
\end{equation}
Here we solve equation~\eqref{eq:ngmass7:system} for the two-dimensional case, which means that all the body trajectories lie in the $xOy$ plane. 

\begin{wrapfigure}{r}{0.36\textwidth}
	\vspace*{-7mm}
		\centering
	\includegraphics[scale=0.25]{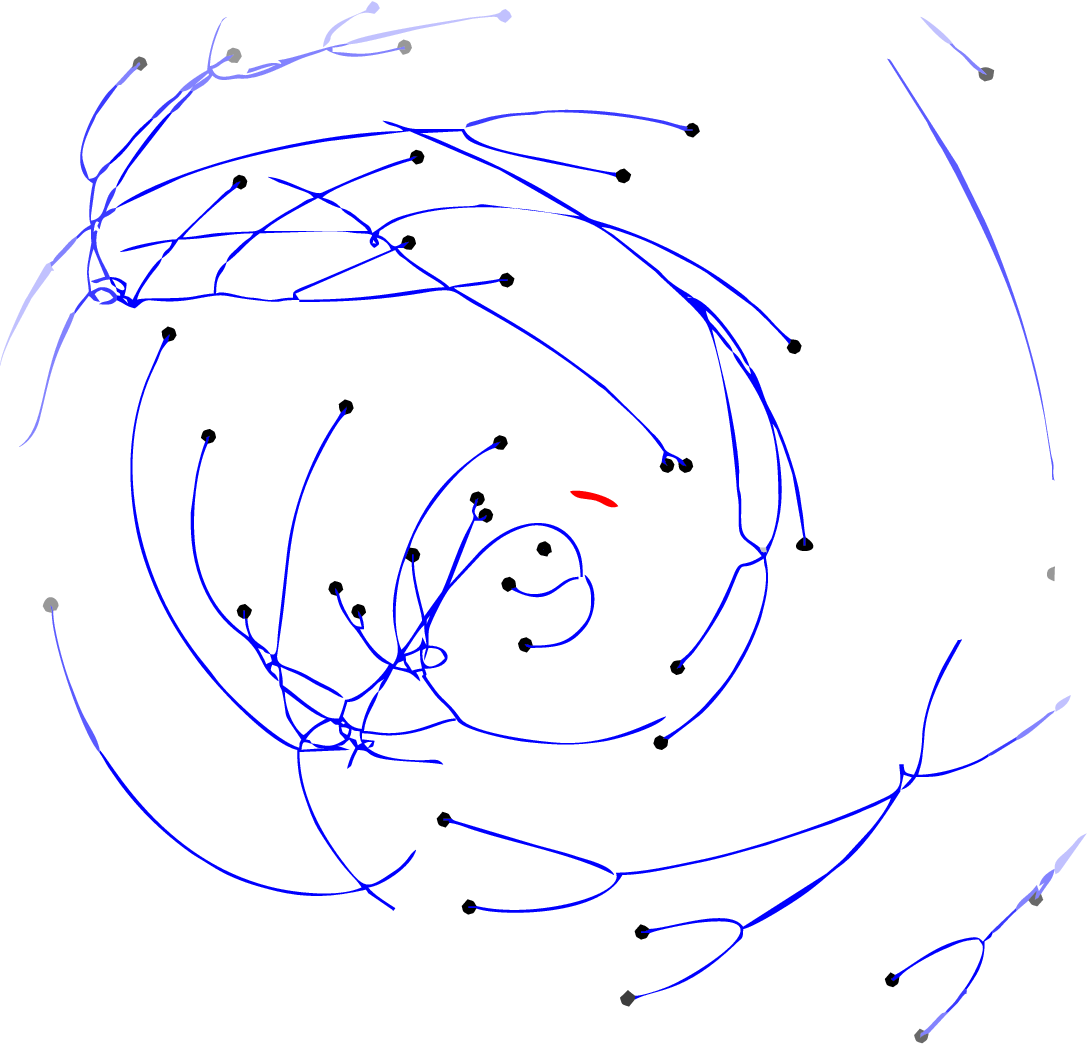}
	\caption{%
		Calculated trajectories of the bodies before and after an absolutely inelastic collision. 
		Joint rotation of the bodies is counterclockwise%
	}
	\label{fig:ngmass7-fig-tracers-accretion}
\end{wrapfigure}
The $i$-th body coordinates are defined by radius $r_i$ and angle~$\lambda_i$. 
The angle is measured counterclockwise from axis~$Ox$. Angle~$\lambda_i$ will be chosen for each body randomly within the range of $0$ to $2\pi$. Initial distribution of the orbit radii will be found via formula~\eqref{eq:ngmass7-alpha-r} with parameters $\alpha=0.97$ and $r_{min}=500\SIm$. Using~\eqref{eq:ngmass7-vm}, let us calculate initial circular velocities~$v_i$ for each cluster body.

Now, as initial conditions for the set of second-order differential equations~\eqref{eq:ngmass7:system} are defined, let us solve the Cauchy problem continuously checking the inter-body distance in order to detect the moment when the collision conditions are fulfilled. Two bodies (spheres) will be regarded as collided if the distance between their centers of mass is shorter than or equal to the semi-sum of their diameters. Since we consider absolutely inelastic collisions, the two bodies continue moving after contacting as an integral whole with the same velocity and in the same direction.

\begin{wrapfigure}{l}{0.45\textwidth}
	\vspace*{-5mm}
	\centering
	\includegraphics[scale=0.26]{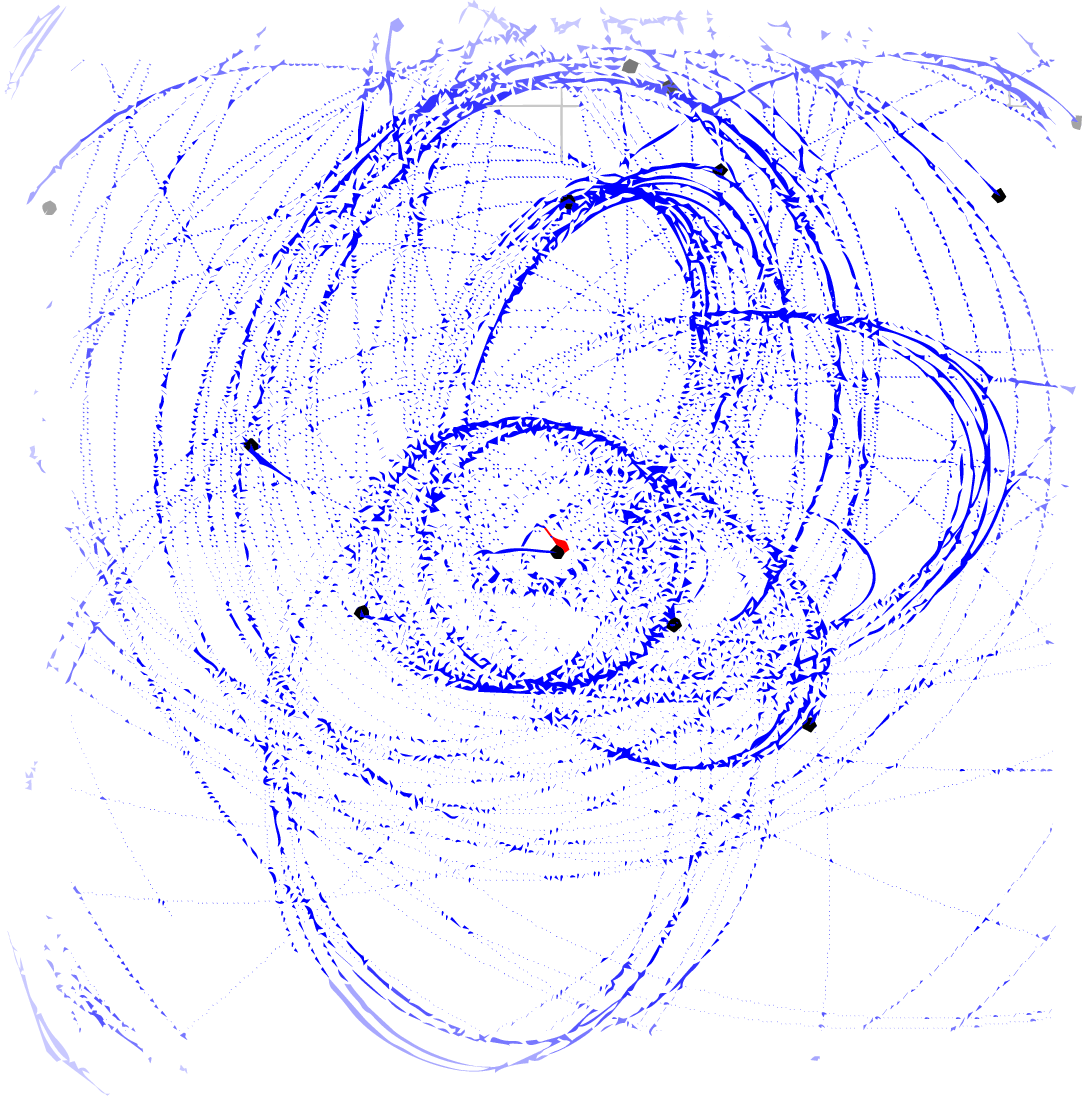}
	\caption{%
		Examples of design trajectories of the cluster gravitating bodies%
	}
	\label{fig:ngmass7-fig-tracers}
\end{wrapfigure}
Fig.~\ref{fig:ngmass7-fig-tracers-accretion} presents the calculated trajectories of the bodies before and after the absolutely inelastic collision, while Fig.~\ref{fig:ngmass7-fig-tracers} presents fragments of a more general trajectory pattern for the gravitating bodies of the model cluster.

\begin{figure}[ht!]
	\centering
	\includegraphics[scale=1]{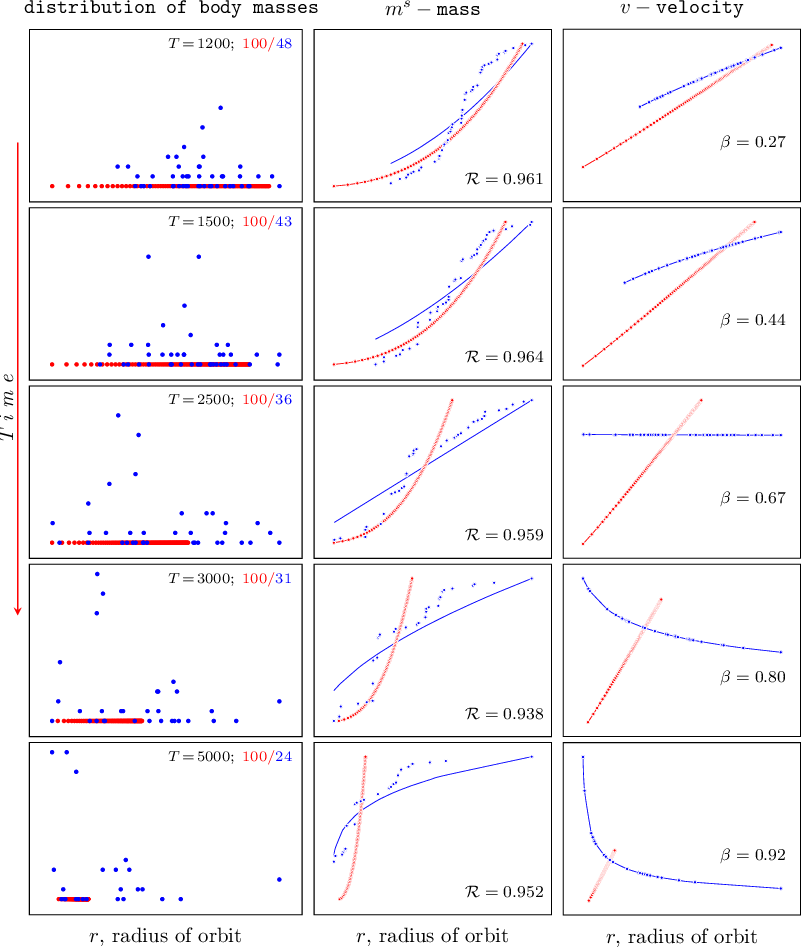}
	\caption{%
		Characteristics of the cluster of gravitating bodies at different evolution durations. Red color indicates the initial state of the cluster bodies, blue color is for the current state. $T$ is the evolution duration, days; $\mathcal{R}$ is the correlation factor; $\beta$ is the evolution number. $\textcolor{red}{100}/\textcolor{blue}{48}$ is the ratio between the initial number of bodies and the remaining one%
	}
	\label{fig:ngmass7-fig-rmv}
\end{figure}

The result of simulating the cluster bodies dynamics in the evolution processes of different durations 
\mbox{$T=$ \small $\{1200; 1500; 2500; 3000; 5000\}\:days$} is given in Fig.~\ref{fig:ngmass7-fig-rmv}.
Red color in the figure indicates the initial state of the system bodies with their initial characteristics, while blue color indicates the system state at time point~$T$.

$\bullet$
The first column (top to bottom) reflects variations in the distribution of body orbits and masses with the cluster evolution duration. Digits in the top-right corners of the first column plots present the ratios between the initial number of bodies and the current one.

$\bullet$
The second column shows how the sequence of masses of the nested \textit{Spheres} changes.
Blue points represent the design distribution of the nested \textit{Spheres’} masses, while the solid blue line is the distribution approximation with the power function. The quality of the~$m^s$ sequence approximation is characterized by correlation factor~$\mathcal{R}$. The cluster mass remains constant during the entire evolution.

$\bullet$
The third column demonstrates the variations in the rotation curve shape and evolution number~$\beta$ during the cluster evolution.


%% file: ngmass7-conclusion-english.tex
%

\section{Conclusions}
\label{section:ngmass7-conclusion}

Numerical simulation of evolution of a gravitating bodies cluster has been performed taking into account gravitational accretion. The simulation has shown that the horizontal section of the cluster rotation curve is defined by the current spatial redistribution of bodies whose total mass remains invariant during the entire evolution. Based on this it is possible to conclude that the nature of the “mysterious” plateau in the galaxy rotation curves~(Fig.~\ref{fig:rotation7-fig-Galaxies-velocity}) may be fully explained by the classical mechanics laws without involving any additional gravitating masses.

The numerical experiment has shown that evolution number $\beta$ of the model cluster is comparable with evolution numbers of real galaxies whose rotation curves contain plateaus~(Table~\ref{tab:ngmass7:beta}).
\begin{table}[h!]
	\centering
	\footnotesize
	\topcaption{%
		Classification of galaxies by the quantitative indicator $\beta$. Assignment of evolution number~$\beta$ to three characteristic types of the galaxy rotation curves~(Fig.~\ref{fig:rotation7-fig-Galaxies-velocity}).%
	}
	\label{tab:ngmass7:beta}
	\begin{tabular}{c|c|c}
		Initial stage & Borderline stage & Final stage \\
		\hline
		\includegraphics[scale=0.7]{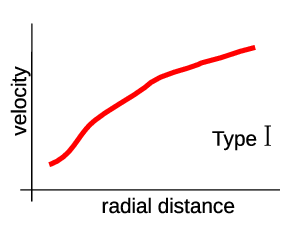} &
		\includegraphics[scale=0.7]{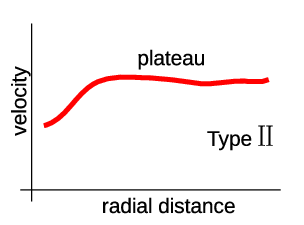} &
		\includegraphics[scale=0.7]{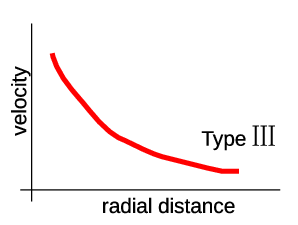} \\ \vspace*{-4mm}
		& & \\
		\input{ngmass7-tab-beta-stage1-english} &
		\input{ngmass7-tab-beta-stage2-english} &
		\input{ngmass7-tab-beta-stage3-english} \\ \vspace*{-3mm}
		& & \\ \hline
	\end{tabular}
\end{table}
Approximate equality of the evolution numbers $\beta$ given in Table~\ref{tab:ngmass7:beta} suggests that the evolution number is invariant to the cluster sizes and masses, which in its turn allows using this parameter as a criterion for galaxies classification with respect to their evolution statuses.

\begin{figure}[h!]
	\centering
	\footnotesize
	\includegraphics[scale=1.12]{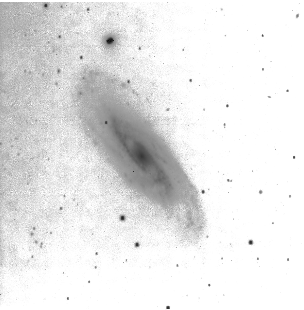}
	\includegraphics[scale=1.3]{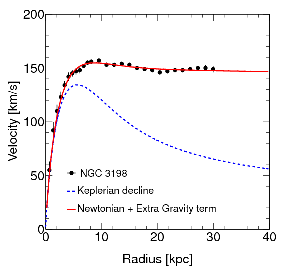}
	\caption{%
		Distribution of the observed and calculated linear velocities of the Galaxy~\texttt{NGC3198} stars~\cite{2013arXiv1312.4528Y} over their distances from the center of rotation%
	}
	\label{fig:NGC3198-neg-velocity}
\end{figure}

\paragraph{Now some words about the “dark matter”.} Due to various force majeure circumstances, the scientific community decided to use in explaining physical nature of the galaxy rotation curve plateau the concept of “dark matter” that is a specific gravitating matter invisible in the electromagnetic range; it is mechanically all-pervading and able to “muffle up” the galaxies and affect their dynamics. Later the refusal of attempts to materialistically interpret the rotation curve plateau gave rise to a great number of metaphysical hypotheses. This entity was used to replenish the allegedly lacking galaxy gravitating mass so as to bring the theoretical rotation curve in compliance with the instrumentally observable one (see Fig.~\ref{fig:NGC3198-neg-velocity}).
The dashed line represents the expected theoretical rotation curve of the galaxy calculated based on the classical mechanics laws. The galaxy mass is estimated by the Doppler measurements of radial velocity of the star outermost from the center of rotation. The theoretical curve (dashed line) demonstrates an unsatisfactory result. This led to a conclusion that the galaxy lacks significantly the gravitating mass and, thus, to the idea of “dark matter” that was assumed to be a source of additional non-observable gravitating matter.

To our opinion, the difference between the rotation curves (Fig.~\ref{fig:NGC3198-neg-velocity}) was caused by two circumstances, namely, by absolute credence to the empirical \textit{mass-luminosity relationship}%
\footnote{%
	Interrelation between the star mass and luminosity was revealed by Jacob Karl Ernst Halm and Arthur Stanley Eddington.%
},
and also by limited capacities of the observational equipment designed to detect faint objects.
\medskip

\textit{%
Assume that material bodies (spheres) involved in the above-described numerical experiment are painted white and black. Let us agree that, contrary to white spheres, black spheres are invisible in the electromagnetic range. This does not affect the algorithm of solution. Such a procedure for simulating the cluster evolution gave a plateau in the cluster rotation curve. Relying on exclusively visual observations, we can detect only white spheres and estimate their masses through their luminosity. Their total mass is evidently insufficient to compose the assumed total mass of the cluster since, as black spheres are invisible, they may be thought to be absent. Extra mass is needed. Here are two ways: to improve the sensitivity of instruments detecting black spheres or to introduce some additional metaphysical gravitating mass. Imaginary painting of the spheres showed that the rotation curve plateau is caused only by the classical matter and classical interpretation of gravitational interaction between material bodies, while the visual absence of material dark spheres is due to poor sensitivity of the instruments.}
\medskip

All the above has led us to the following conclusion: the plateau in the galaxy rotation curve is the reflection of dynamics of all the gravitating matter forming the galaxy. The additional gravitating mass is not needed regardless of its interpretation.


%% file: ngmass7-tab-beta-stage1-english.tex
%

\begin{tabular}[t]{r|c}
	Cluster                   & $\beta$\\ 
	\hline
	\vspace*{-3mm}      &\\
	$NGC598 (M33)$    & 0.496  \\
	$NGC4010$            & 0.406  \\
\end{tabular}


%% file: ngmass7-tab-beta-stage2-english.tex
%

\begin{tabular}[t]{r|c}
	Cluster             & $\beta$\\
	\hline 
	\vspace*{-3mm}&\\
	$NGC6503$               & 0.660  \\
	$NGC3198$               & 0.604  \\
	$NGC2403$               & 0.548  \\
	$NGC7331$               & 0.675  \\
	$NGC4217$               & 0.642  \\
	$UGC6917$               & 0.513  \\
	Milky Way                 & 0.678  \\
	\textcolor{blue}{Computational model} & \textcolor{blue}{0.673}  \\
\end{tabular}



%% file: ngmass7-tab-beta-stage3-english.tex
%

\begin{tabular}[t]{r|c}
	Cluster             & $\beta$\\
	\hline
	\vspace*{-3mm}&\\
	Solar system        & 0.998  \\
	$NGC4138$           & 0.728  \\
\end{tabular}
